\documentclass[twocolumn]{aastex63}
\usepackage{amsmath, amstext}
\usepackage[T1]{fontenc}
\usepackage{apjfonts}
\usepackage[figure, figure*]{hypcap}

\def\lsim{\mathrel{\raise.3ex\hbox{$<$\kern-.75em\lower1ex\hbox{$\sim$}}}}
\def\gsim{\mathrel{\raise.3ex\hbox{$>$\kern-.75em\lower1ex\hbox{$\sim$}}}}
\def\gtwid{\mathrel{\raise.3ex\hbox{$>$\kern-.75em\lower1ex\hbox{$\sim$}}}}
\def\proptwid{\mathrel{\raise.3ex\hbox{$\propto$\kern-.75em\lower1ex\hbox{$\sim$}}}}

\newcommand{\ud}{\mathrm{d}}
\newcommand{\bhspin}{a_*}

\newcommand{\OO}[1]{}

\shortauthors{Palumbo, Wong, and Prather}

\defcitealias{PaperI}{EHTC~I}
\defcitealias{PaperII}{EHTC~II}
\defcitealias{PaperIII}{EHTC~III}
\defcitealias{PaperIV}{EHTC~IV}
\defcitealias{PaperV}{EHTC~V}
\defcitealias{PaperVI}{EHTC~VI}

\begin{document}

\title{Discriminating Accretion States via Rotational Symmetry in Simulated Polarimetric Images of M87}

\correspondingauthor{Daniel~C.~M.~Palumbo}
\email{daniel.palumbo@cfa.harvard.edu}

\author[0000-0002-7179-3816]{Daniel~C.~M.~Palumbo}
\affil{Center for Astrophysics $|$ Harvard \& Smithsonian, 60 Garden St., Cambridge, MA 02138, USA}
\affiliation{Black Hole Initiative at Harvard University, 20 Garden St., Cambridge, MA 02138, USA}
\author[0000-0001-6952-2147]{George~N.~Wong}
\affil{Department of Physics, University of Illinois, 1110 West Green St., Urbana, IL 61801, USA}
\affil{CCS-2, Los Alamos National Laboratory, P.O. Box 1663, Los Alamos, NM 87545, USA}
\author[0000-0002-0393-7734]{Ben S. Prather}
\affil{Department of Physics, University of Illinois, 1110 West Green St., Urbana, IL 61801, USA}

\begin{abstract}

In April 2017, the Event Horizon Telescope observed the shadow of the supermassive black hole at the core of the elliptical galaxy Messier 87. While the original image was constructed from measurements of the total intensity, full polarimetric data were also collected, and linear polarimetric images are expected in the near future. We propose a modal image decomposition of the linear polarization field into basis functions with varying azimuthal dependence of the electric vector position angle. We apply this decomposition to images of ray traced general relativistic magnetohydrodynamics simulations of the Messier 87 accretion disk. For simulated images that are physically consistent with previous observations, the magnitude of the coefficient associated with rotational symmetry, $\beta_2$, is a useful discriminator between accretion states. We find that at 20 $\mu$as resolution, $|\beta_2|$ is greater than 0.2 only for models of disks with horizon-scale magnetic pressures large enough to disrupt steady accretion. We also find that images with a more radially directed electric vector position angle correspond to models with higher black hole spin. Our analysis demonstrates the utility of the proposed decomposition as a diagnostic framework to improve constraints on theoretical models.

\end{abstract}


\section{Introduction} \label{sec:intro}

In April 2017, the Event Horizon Telescope (EHT) observed the compact core of the nearby giant elliptical galaxy Messier 87 at horizon scale resolution \citep[][hereafter EHTC~I-VI]{PaperI,PaperII,PaperIII,PaperIV,PaperV,PaperVI}.
The reconstructed images revealed a distinct central brightness depression consistent with the shadow of a black hole \citepalias{PaperIV, PaperV, PaperVI}. To identify which black hole system parameters are consistent with the data, a large library of ray traced general relativistic magnetohydrodynamic (GRMHD) images of simulated accretion flows was generated. Hereafter, we refer to these images as the image library. Though some models were excluded by the first EHT results, the majority were found to be consistent with the total intensity data \citepalias{PaperV}.  

The accreting material around the central black hole in Messier 87 (hereafter M87) is typically modeled as a radiatively inefficient accretion flow forming a geometrically thick disk of infalling plasma \citep{Ichimaru_1977, Rees_1982,Narayan_1994, Narayan_1995,Reynolds_1996}. At M87-like mass and accretion rates, radiation at the 230 GHz EHT operational frequency is dominated by synchrotron emission
\citep[see, e.g.,][]{Yuan_Narayan_2014}. In the synchrotron process, electrons are confined to move in helical orbits about magnetic field lines. 
This motion sets a characteristic orientation for the electromagnetic fields that are produced and results in a polarization perpendicular to the orientation of the magnetic field lines. \deleted{In the limit of weak internal Faraday rotation, the observed linear polarization can be used to probe the structure of the local magnetic field.}

Accretion flows can be divided into two qualitatively different states according to the properties of their steady-state magnetic fields. In the magnetically arrested disk (MAD) state, the magnetic pressure in the disk near the horizon is large enough to counterbalance the inward ram pressure of the flow \citep{Ichimaru1977,Igumenschchev_2003,Narayan2003}.
MAD flows are characterized by energetic, quick, violent accretion events and often have higher accretion efficiencies compared to their standard and normal evolution (SANE) counterparts \added{\citep{Narayan_2012}}.

Since the structure and strength of the magnetic field at the horizon parametrizes the black hole accretion flow state space, linear polarization data may be an efficient discriminator among the underlying models. The 2017 EHT observations contain full polarimetric information and thus the properties of M87's magnetic field configuration may be distinguishable by analysis of polarimetric images.

\added{Very long baseline interferometry (VLBI) observations have constrained the linear fractional polarization of the M87 core to less than $0.07\%$ \citep{homan_2006}. More recent observations show fractional polarization up to 4\% near the core of M87 \citep{Walker_2018}. \citet{Hada_2016} finds  fractional polarization up to 20\% in regions downstream from the core. These ordered features suggest structured magnetic field in central regions sampled by the EHT. \citet{kuo_2014} observed M87 with the Submillimeter array at 230 GHz to derive a rotation measure (RM) between $-7.7\times10^5$ and $3.4\times10^5$ rad ${\rm m}^{-2}$. Though this measurement is highly uncertain, tighter constraints are expected from forthcoming EHT results.} Prior work has linked linear polarization fraction and rotation measure in emission from accretion flows to magnetic field structure in simulations of M87 \citep{Broderick_2009, Mosci_faraday_2017}.

In this article, we present a framework for evaluating rotational coherence in the linear polarimetric fields of arbitrary images. We use this framework to provide a focused analysis of nearly face-on accretion flow images in a simulated library of M87-like images. We generate a discriminator between the MAD and SANE accretion states and identify trends in the coefficients with respect to black hole spin. The differences in polarization structure across the image library support a path to inference of the magnetic field structure that may be applicable to EHT polarimetry of M87.

This paper is structured as follows. In Section~\ref{sec:decomposition} we define our decomposition procedure and motivate its application with several simple examples. In Section~\ref{sec:imagelibrarydiscr} we apply our decomposition procedure to a set of physically motivated simulated images and detail some observed statistical trends. We conclude with a brief survey of model limitations and provide a discussion of future directions and applications in Section~\ref{sec:discussion}.

\section{Decomposition of Linear Polarization}
\label{sec:decomposition}

Images of face-on black hole accretion disks exhibit a ring-like structure that aligns with the symmetry axis of the Kerr spacetime. Although this symmetry persists in the magnitude of polarized intensities, the map of the electric vector position angle (EVPA) depends strongly on the orientation and strength of magnetic fields in the flow. 
The higher magnetic field strengths present in MADs may lead to increased azimuthal symmetry in EVPA, motivating a symmetry-based decomposition of linearly polarized images to distinguish MAD and SANE states. We describe our method below.

\begin{figure*}[t]
    \centering
    \includegraphics[width=.99\textwidth]{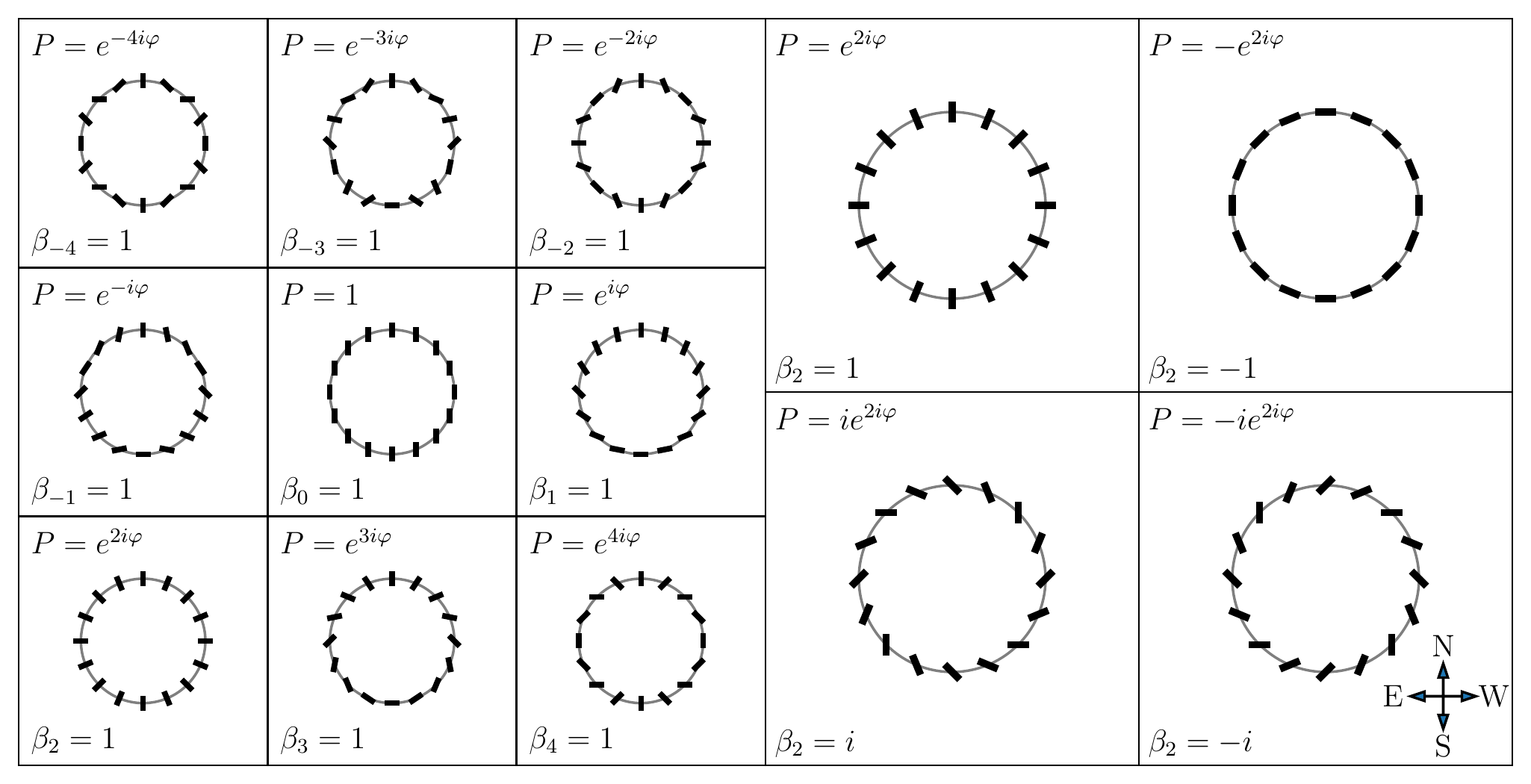}
    \caption{Left grid: examples of the electric vector position angle for periodic polarization fields plotted along a ring of unit radius, along with corresponding $\beta_m$ values for $-4 \leq m \leq 4$. Polarization fields are chosen to produce positive real values of $\beta_m$, which correspond to vertical electric vector position angle at the top of the image. Right grid: Same as left, but showing only the rotationally symmetric $m=2$ mode with four phases in $\beta_2$. }
    \label{fig:beta_2_ringplot}
\end{figure*}

\begin{figure*}[t]
    \centering
    \includegraphics[width= \textwidth]{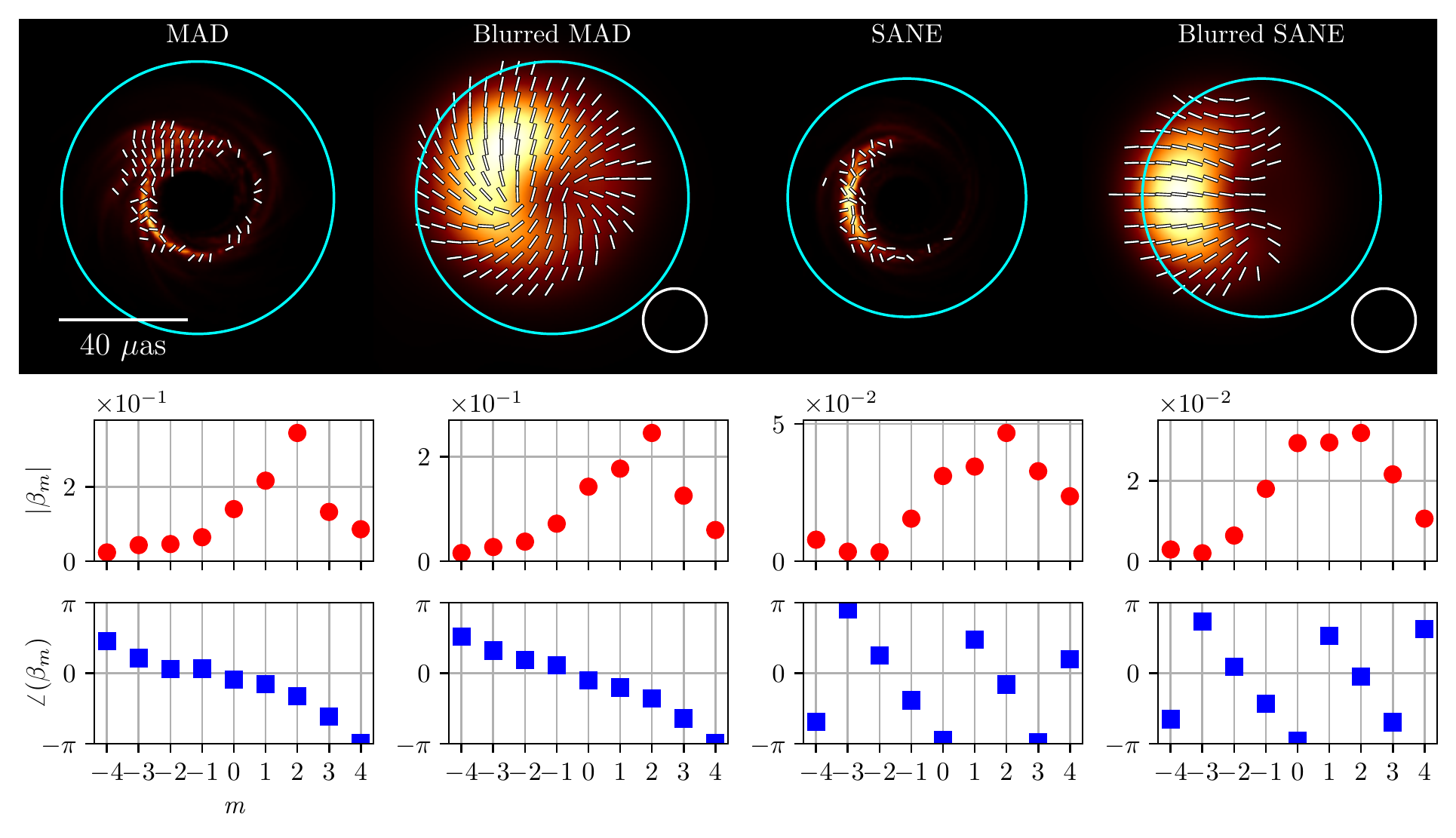}
    \caption{Comparison of linear polarimetric decomposition of example MAD and SANE images consistent with observational criteria and approximately equal total flux. \added{Each model has $a_* = 0.94$ and $R_{\rm high}=80$.} The decomposition is applied within the annulus stretching from the blurred \texttt{rex}-fit ring diameter to twice its half-width in each direction, centered at the \texttt{rex}-fit ring center. Color shows unpolarized Stokes $I$ intensity normalized to unity and ticks show EVPA. Images are shown with and without a 20 $\mu$as blurring kernel applied to all Stokes grids. White circles show the blurring kernel; the blue circle shows the outer edge of the \texttt{rex} annulus, while the inner edge extends to zero. Both images are of M87-like simulations with a $6.2 \times 10^9$ ${\rm M}_\odot$ central black hole of spin $\bhspin = 0.94$, viewed at $17^\circ$ inclination to the black hole spin axis, and with identical models for electron temperature.
    In figures, EVPA tick marks are shown where both fractional polarization exceeds 1\% and Stokes $I$ intensity exceeds 10\% of its maximum value. The MAD snapshot is dominated by power in the $m=2$ mode. Both blurring and integrating over larger scales impose coherence, decreasing power in higher modes.}
    \label{fig:mad_v_sane_example}
\end{figure*}

\subsection{Decomposition Definition}

For our analysis, we take advantage of the inherent ring-like structure present in black hole images at low inclination by working in polar coordinates $\left(\rho, \varphi\right)$ where the radial distance $\rho$ is measured from the image center and the azimuthal angle $\varphi$ is measured east of north on the sky. We express the linear polarization on the image in terms of the complex-valued polarization field $P(\rho, \varphi) \equiv Q(\rho, \varphi) + i U(\rho, \varphi)$ where $Q$ and $U$ are the usual Stokes intensities. The corresponding EVPA $\chi$ is measured east of north on the sky and can be written in terms of the complex phase of the polarization field $\angle(P)$:
\begin{align}
    \chi &= \dfrac{1}{2}\arctan{\frac{U}{Q}} = \dfrac{1}{2}\angle(P).
\end{align}

We project each image onto a set of basis functions defined in the polarization domain as $P_m(\varphi) \equiv e^{i m \varphi}$ to pick out particular modes of azimuthal symmetry. While orthogonal, our basis functions are not complete over the polarimetric image domain because they cannot reproduce radial structure and do not contain absolute polarized flux information without an accompanying Stokes $I$ image. This is intentional: rather than merely reproduce the polarization map, we wish our decomposition coefficients to be a measure of coherent polarization for particular azimuthal angular dependencies about the image center. Further, because fiducial model images exhibit a sharp ring-like structure, condensing the data along the radial dimension is not expected to lead to significant information loss. 

We define the decomposition coefficient $\beta_m$ to be the scalar product between the basis image and the $P$ image, restricted to an annulus:

\begin{align}
    \beta_m &= \dfrac{1}{ I_{\rm ann}} \int\limits_{\rho_{\rm min}}^{\rho_{\rm max}} \int\limits_0^{2 \pi} P(\rho, \varphi)P_m^* (\varphi) \; \rho \mathop{\ud\varphi}  \mathop{\ud\rho}\nonumber\\
    &= \dfrac{1}{I_{\rm ann}} \int\limits_{\rho_{\rm min}}^{\rho_{\rm max}} \int\limits_0^{2 \pi} P(\rho, \varphi) \, e^{- i m \varphi} \; \rho \mathop{\ud\varphi}  \mathop{\ud\rho},\\
    I_{\rm ann} &= \int\limits_{\rho_{\rm min}}^{\rho_{\rm max}} \int\limits_0^{2 \pi} I(\rho, \varphi) \; \rho \mathop{\ud\varphi} \mathop{\ud\rho}.
\end{align}
We have normalized by $I_{\mathrm{ann}}$, the total Stokes $I$ flux in the annulus, and $\rho_{\mathrm{min}}$ and $\rho_{\mathrm{max}}$ set the radial extent of the annulus.

For each image, we identify the image center and the radial extent of the annulus according to the ring extractor (\texttt{rex}) procedure described in detail in Section 9 of \citetalias{PaperIV}. \texttt{rex} identifies the ring center as the point which is most equidistant from peak emission along 360 azimuthal slices. The ring width is taken to be the mean full width of half-maximum of the emissivity evaluated along each azimuthal angle from the central point.

For this analysis, we take $\rho_{\mathrm{max}} - \rho_{\mathrm{min}}$ to be twice the ring width reported by the \texttt{rex}-fit ring profile. For GRMHD snapshots blurred to nominal imaging resolutions, this choice should ensure that the decomposition coefficients are well-behaved under perturbations to annulus shape and position. Because the nominal image resolution sets the minimum feature size, the area integral will necessarily include at least two distinct resolution elements along any given radius thereby enabling a meaningful and consistent measure of radial coherence in the polarized image. 

Each $\beta_m$ coefficient is a dimensionless complex number with magnitude corresponding to the amount of coherent power in the $m^{\mathrm{th}}$ mode and with phase corresponding to the average pointwise rotation of the image polarization relative to a fiducial EVPA orientation which we define to be vertical along the $\varphi = 0$ image axis. The process can also be thought of as a radially averaged azimuthal Fourier transform of the complex polarization field where the $\beta_m$ coefficients are Fourier coefficients corresponding to the integral Fourier modes. 

If the jet in M87 is aligned with the spin axis of the central black hole as is believed, then it is expected that our viewing angle to the accretion system should be small \citep[see, e.g.,][]{WangZhou_2009,MLWH_2016,Walker_2018}. In such nearly face-on systems, the $m=2$ mode is of particular interest because of the expected axisymmetric structure in horizon-scale images.

We present a few trivial examples of ring-valued linear polarization fields corresponding to $-4 \leq m \leq 4$ periodic modes with different phases in Figure~\ref{fig:beta_2_ringplot}. Note that the $\beta_2$ coefficient projects out a polarimetric symmetry akin to the $E$ and $B$ modes typically used in studies of polarization in the cosmic microwave background \citep[e.g.,][]{Kamionkowski_2016}. In this formalism, the real part of the $\beta_2$ coefficient corresponds to $E$ and the imaginary part corresponds to $B$.

In \autoref{fig:mad_v_sane_example}, we provide an example of the decomposition applied to images of models that pass \citetalias{PaperV} observational criteria. This example illustrates two strong features of the decomposition. First, blurring increases coherence in polarimetric structure and reduces power in modes with higher $|m|$. Second, the $m=2$ rotationally symmetric mode can be dominant in both MAD and SANE model snapshots, though in this case, the MAD snapshot has a much larger $m=2$ coefficient. Further, the complex phase of the $\beta_2$ component $\angle(\beta_2)$ encodes the dominant direction of the EVPA spiral. In this MAD snapshot, $\angle(\beta_2) \sim- \pi/2$, corresponding to the EVPA in the bottom right of \autoref{fig:beta_2_ringplot}, while the SANE snapshot has $\angle(\beta_2)\sim 0$, corresponding to a radially directed EVPA.

\subsection{Interferometric Signatures of Rotationally Symmetric Polarization}

\begin{figure*}[t]
    \centering
    \includegraphics[width= \textwidth]{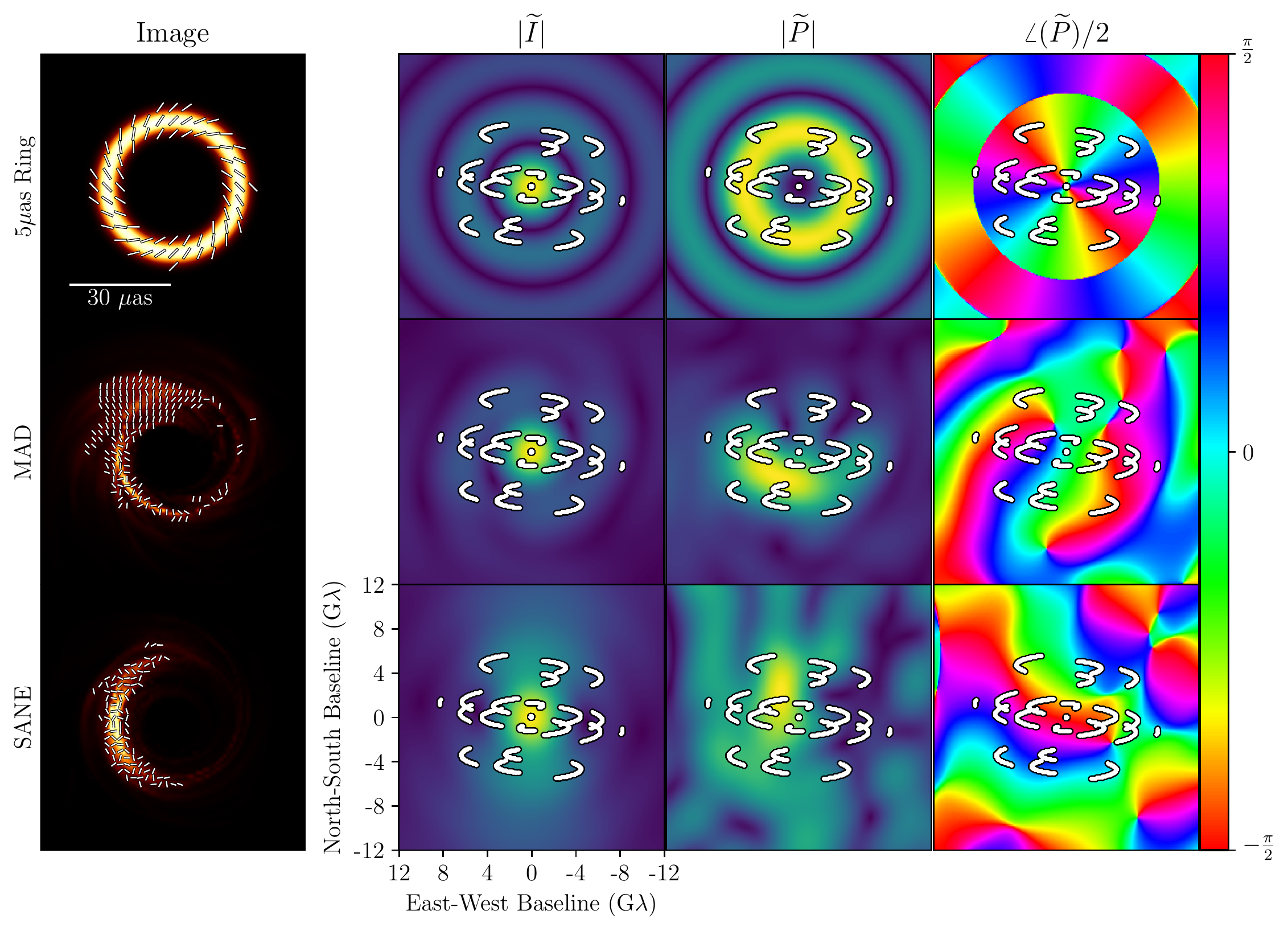}
    \caption{Visibility amplitudes $|\widetilde{I}|$ and $|\widetilde{P}|$, and $\frac{1}{2}\angle(\widetilde{P})$ for three example images. The top row presents a 40$\mu$as diameter boxcar ring of width $1 \mu$as blurred by a 5 $\mu$as Gaussian kernel \added{with constant fractional polarization corresponding to $\beta_2 = -i$}. The bottom two correspond to the GRMHD examples presented in \autoref{fig:mad_v_sane_example}. EHT 2017 baseline coverage from April 11 is overlaid. Rotational symmetry in the EVPA transfers to the visibility domain, but rotates by $90^\circ$ on short baselines due to the factor of $- \beta_2= i$ found in \autoref{eq:pvis_example}. Visibility amplitudes are normalized and shown in linear scale.}
    \label{fig:mad_v_sane_fourier}
\end{figure*}

\begin{figure*}
    \centering
    \includegraphics[width=\textwidth]{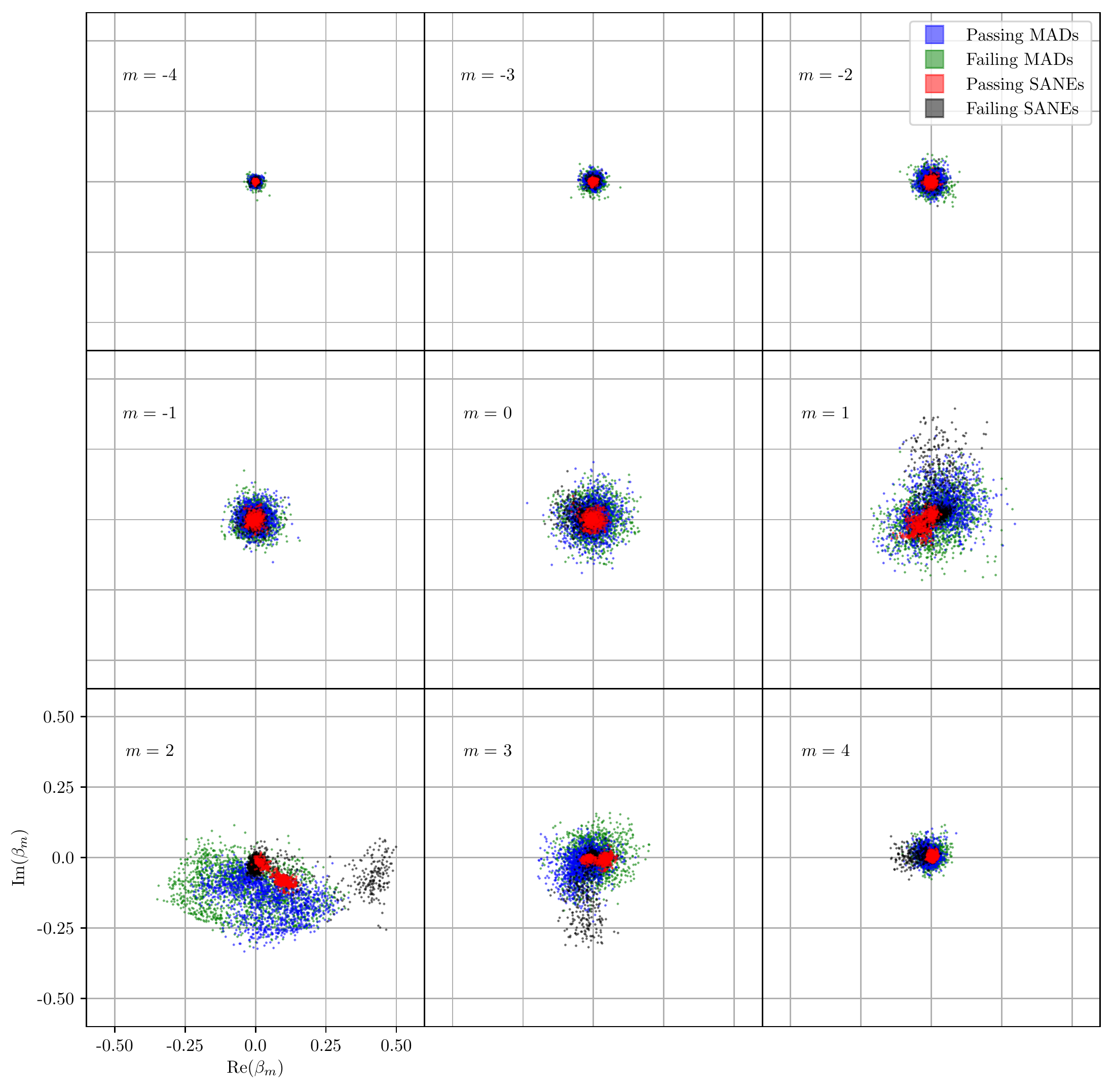}
    \caption{Complex $\beta_m$ coefficients with $-4 \leq m \leq 4$ for the fiducial ray tracing parameters of the GRMHD library after blurring with a 20 $\mu$as beam. Coefficients are normalized by the Stokes $I$ annular flux after integrating over a region set by the \texttt{rex}-fit radius and width $\rho_{\rm \texttt{rex}} \pm w_{\rm \texttt{rex}}$. Models that are not self-consistent or that are ruled out by prior observational constraints are labeled as failing.}
    \label{fig:complex_all}
\end{figure*}

An interferometer measures visibilities $\widetilde{I}(u, \psi)$ of an image $I(\rho,\varphi)$ according to \citep[see, e.g.,][]{TMS}
\begin{align}
    \widetilde{I}(u, \psi)&=\iint I(\rho, \varphi)\,e^{-i 2 \pi \rho u \cos(\psi-\varphi)} \; \rho\, \mathop{\ud \rho} \mathop{\ud \varphi},
\end{align}
where $u$ and $\psi$ give the location of the interferometric baseline when projected onto the image plane perpendicular to the line of sight. Here, $u$ is the magnitude of the baseline vector and $\psi$ is measured east of the positive $u_y$ axis. This definition holds for all Stokes parameters. We adopt the notation $\widetilde{I}$, $\widetilde{Q}$, and $\widetilde{U}$ for the visibilities associated with Stokes $I$, $Q$, and $U$ respectively. The linear polarization visibilities $\widetilde{P}$ are then $\widetilde{Q}+i \widetilde{U}$ and have interferometric EVPA given by $\frac{1}{2}\angle(\widetilde{P})$.

We now identify the $\widetilde{P}$ signatures corresponding to azimuthal symmetry in $P$.
Consider the simple case of a rotationally symmetric Stokes $I$ image with constant fractional polarization $p$. If the polarization evolves azimuthally according to a single mode $m$, then
\begin{align}
    I(\rho, \varphi) &\equiv I(\rho),\\
    P(\rho, \varphi) &\equiv p I(\rho) e^{i m \varphi}
\end{align}
and thus $P(\rho, \varphi)$ is separable in $\rho$ and $\varphi$. The polarized visibilities can then be written
\begin{align}
    \widetilde{P}(u, \psi)&=p\int\limits_0^\infty  I(\rho) \left[ \int\limits_0^{2 \pi}e^{i m \varphi} e^{-i 2 \pi  \rho u \cos(\psi-\varphi)} \mathop{\ud \varphi}\right] \rho \mathop{\ud \rho} .
    \label{eq:pvis}
\end{align}

The integral in $\varphi$ produces Bessel functions of the first kind $J_m$ and leaves the azimuthal structure intact up to a phase dependence that is determined by $m$ and the sign of $J_m$:
\begin{align}
    \int\limits_0^{2 \pi}e^{i m \varphi} e^{-i 2 \pi \rho u \,\cos(\psi-\varphi)} \mathop{\ud\varphi} &= 2 \pi   i^{-m} J_m(2 \pi u \rho)e^{i m \psi}.
\end{align}
Because what remains of \autoref{eq:pvis} is an integral in $\rho$, the angular dependence on $e^{i m \psi}$ will be present in the visibility domain and thus dependence on the image angle $\varphi$ is imprinted on the Fourier domain angle $\psi$. 

The image of a thin ring polarized according to the rotationally symmetric $m=2$ mode is particularly relevant to our analysis. We can fix the EVPA of the $m=2$ mode by setting the phase of $\beta_2$.

A thin polarized ring with diameter $d$ is then given in the image plane by \citep{Johnson_2019}
\begin{align}
    I(\rho,\varphi) &= \dfrac{1}{\pi d}\delta{\left(\rho-\dfrac{d}{2}\right)},\\
    P(\rho, \varphi) &= \beta_2 \,p\dfrac{1}{\pi d}\delta{\left(\rho-\dfrac{d}{2}\right)\, e^{i 2 \varphi}}.
\end{align}

The corresponding visibilities are then
\begin{align}
    \widetilde{I}(u,\psi) &= J_0 (\pi d u),\\
    \widetilde{P}(u,\psi) &=  -\beta_2 p J_2 (\pi d u)\,e^{i 2 \psi}.
    \label{eq:pvis_example}
\end{align}
Evidently, the signature of the $m=2$ polarization mode is two-fold in $\widetilde{P}$, comprising both the $J_2$ Bessel function and the $e^{i 2 \psi}$ azimuthal dependence in the Fourier domain. Each signature is readily identifiable in synthetic data. 

\autoref{fig:mad_v_sane_fourier} illustrates these interferometric polarized signatures for three model images: an analytic \added{$\beta_2 = -i$} ring of diameter $40 \mu$as \replaced{whose profile is}{with a profile} built from a $1 \mu$as rotated boxcar blurred by a $5\mu$as Gaussian, and the MAD and SANE snapshots shown in \autoref{fig:mad_v_sane_example}. The MAD model shows interferometric signatures similar to those of the $\beta_2 = - i$ right-handed spiral, \added{such as the the similar azimuthal evolution of phase and the sudden amplitude decay near the second $J_2$ null}.

\autoref{fig:mad_v_sane_fourier} also shows the overplotted baseline coverage of M87 provided by the 2017 EHT array. Even though the array only provided a sparse sampling of the Fourier plane, we expect that data at the current EHT resolution will still be sensitive to the salient differences in EVPA structure between MAD and SANE models. \added{In particular, the angular half-period of a mode is $ \pi / m$; if the majority of image structure is located at a single diameter $D$, then the spatial scale of this half-period is $\pi D /2 m$. In order for the EHT to resolve a mode, we may demand that the effective resolution be smaller than the half-period. That is, for an angular resolution $\theta_r$, the largest $m$ we might reasonably expect to resolve is constrained by
\begin{align}
    |m_{\rm max}| &\lesssim \frac{\pi D }{2 \theta_r}\\
    &\lesssim \pi \left(\frac{D}{40\mu{\rm as}}\right)\left(\frac{\theta_r}{20\mu{\rm as}}\right)^{-1}.
\end{align}
At the $\sim 20\mu$as resolution of the 230 GHz EHT array, we may reasonably expect to resolve up to $|m|=3$ for M87.}

\section{Image Library Parameter Discrimination}
\label{sec:imagelibrarydiscr}

\begin{figure}
    \centering
    \includegraphics[width=0.48\textwidth]{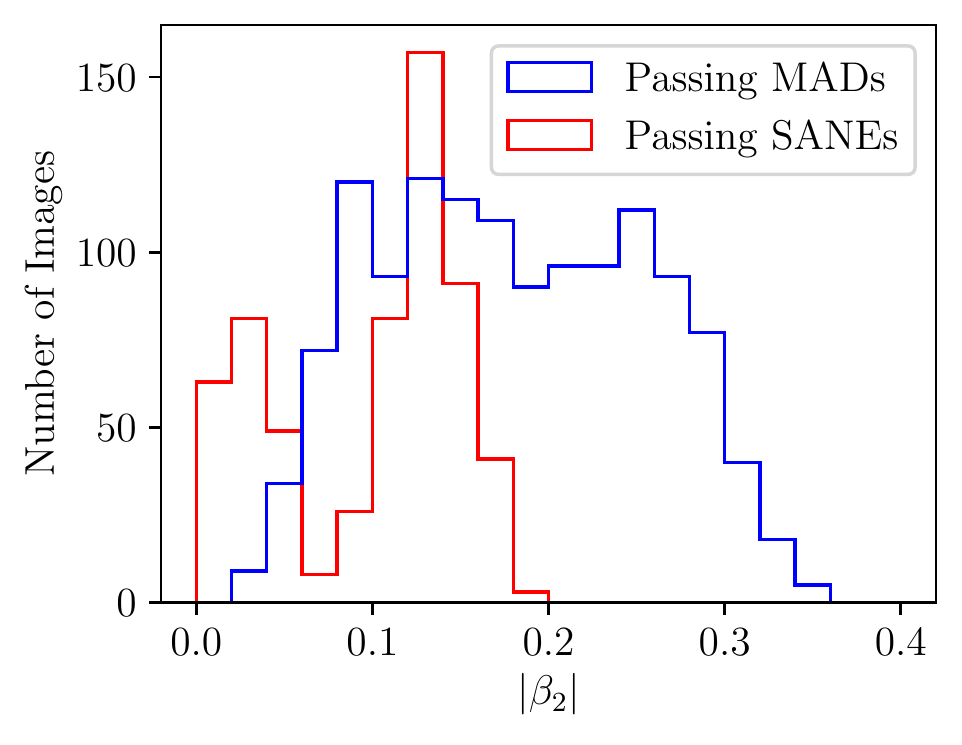}
    \caption{\replaced{Stacked}{Overlaid} histograms of $\beta_2$ coefficient magnitudes shown for MAD and SANE models that are physically consistent with observational criteria.
    No passing SANE model has $|\beta_2|>0.20$.}
    \label{fig:m2_mags}
\end{figure}

\begin{figure*}
    \centering
    \includegraphics[width=\textwidth]{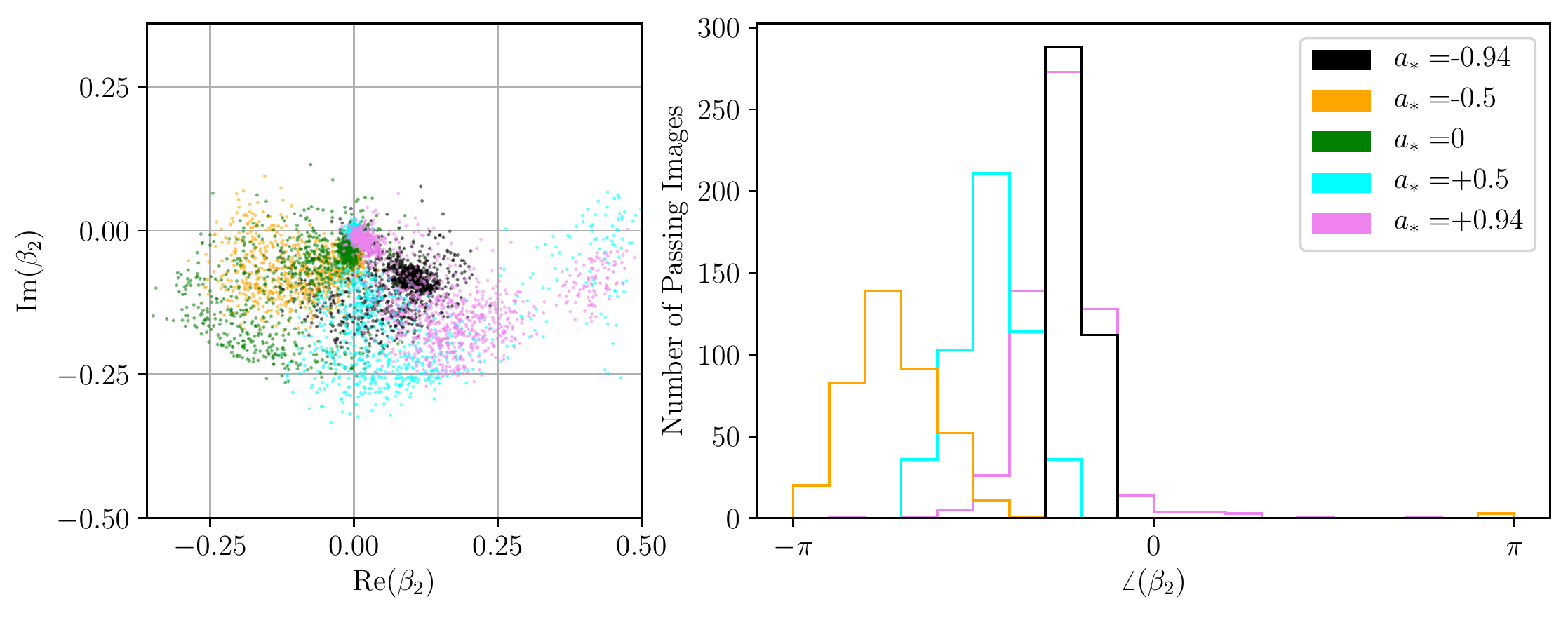}
    \caption{Distribution of the complex $\beta_2$ coefficient (left) and \replaced{stacked}{overlaid} histograms of the phase of $\beta_2$ (right) colored by spin values. Complex coefficients are shown for all models, whereas phases are shown only for passing models. The phase of the $\beta_2$ coefficient reflects increasingly radial EVPA at high spins. As shown in the MAD $\beta_2$ distribution, magnetic field symmetries in left handed flows correspond to right-handed EVPA maps and corresponding $\beta_2$ phases.
    }
    \label{fig:m2_spin}
\end{figure*}

By applying the decomposition described in \autoref{sec:decomposition} to ensembles of simulated images and obtaining representative coefficient distributions, we can identify how changing the physical parameters of the black hole accretion system affects the values of the decomposition coefficients. This analysis provides a new domain which might be used to inform parameter extraction efforts. Because we expect polarimetric images of M87 to be available soon, we apply our decomposition to a set of images generated with M87-like parameters. We present the resulting coefficient distributions and then focus on the $m=2$ coefficient which appears to provide the strongest discriminating power.

\subsection{Image Library Description}

As a part of the analysis effort presented in \citetalias{PaperV}, high-resolution GRMHD simulations of MAD and SANE accretion disks with dimensionless black hole spins $\bhspin \equiv J c / G M^2$ ranging between $-0.95 < \bhspin < 0.99$ were generated.
Here, $J$ is the angular momentum of the black hole, and negative values of $\bhspin$ correspond to anti-parallel black hole and disk angular momentum vectors.
Each GRMHD simulation was used to generate a set of $\geq\,100$ polarimetric images evenly spaced in time via
general relativistic ray tracing. The ray tracing calculation was performed using the fast light approximation and assumed pure synchrotron emission and absorption from a thermal electron distribution.
As part of the analysis, a set of observational constraints and self-consistency checks was applied to each model. A full description of the library and cuts applied to the models is available in \citetalias{PaperV}.

We restrict our analysis to images of black holes with mass $6.2 \times 10^9 M_{\odot}$ and average compact flux densities $F_\nu(230\,\mathrm{GHz}) = 0.5$ Jy. Each image was generated at an inclination of $17^{\circ}$ relative to the angular momentum axis of the black hole in accordance with large-scale estimates of jet inclination \citep[see, e.g.,][]{Walker_2018} and in order to produce the characteristic Stokes $I$ brightness asymmetry on the correct side of the image.
We take images from both MAD and SANE models with dimensionless spins $\pm 0.94, \pm 0.5$, and $0$ with \deleted{six different} electron temperature prescriptions \added{determined by six values of the $R_{\rm high}$ parameter adapted from \citet[][]{Mosci_2016}}. Our fiducial image set is split evenly among each of the image parameters and includes 3000 each of MAD and SANE images. Of these, 1300 MAD and 600 SANE images pass the observation and consistency checks.  The GRMHD simulations we consider were generated with  {\tt{}iharm3D} \citep{Gammie_HARM_2003} and the radiative transfer calculation was performed by {\tt{}ipole} \citep{Moscibrodzka2018}.

\subsection{Parameter Discrimination Results}

We apply the decomposition described in \autoref{sec:decomposition} to each image in the image library to compute $\beta_m$ coefficients for $-10 \le m \le 10$.
Each image is first blurred by a 20 $\mu$as Gaussian kernel.  
The 20 $\mu$as size corresponds to slightly less than the nominal beam of the EHT array; EHT imaging algorithms routinely reconstruct images with super-resolution finer than this scale \citep[see \citetalias{PaperIV};][]{Chael_2016,Kuramochi_2018, Palumbo_2019}. We then simultaneously center the image and measure a ring profile using \texttt{rex}. Finally, we measure coefficients of the blurred images and examine distributions of the $\beta_m$ coefficients for MAD and SANE simulations.

\autoref{fig:complex_all} shows the distribution of the $\left|m \right| \le 4$ coefficients across the selected library images organized by MAD or SANE as well as whether the snapshot belongs to a model that passed the consistency checks. These checks include comparisons to previous observations as well as a self-consistency verification that the numerical models do not produce too much radiation to have been accurately simulated by non-radiative GRMHD.
Though the $m=1$ and $m=3$ modes appear to segregate the MAD and SANE distributions, the largest apparent separation arises in the $m=2$ mode at the bottom left in the figure.

A small subset (less than 10\%) of SANE snapshots exhibit well-ordered polarization fields with even larger $\beta_2$ magnitudes than MAD models. These snapshots correspond to prograde spin SANE models in which the ion and electron temperatures are set equal. This choice results in a shift of emission outwards beyond the inner accretion flow and thus 
EVPA likely traces the magnetic field structure of the disk proper in these models. The original \citetalias{PaperV} analysis uniformly rejected these models.

The distribution of $\beta_2$ magnitudes in \autoref{fig:m2_mags} suggests that discrimination between MAD and SANE is tractable even if the EVPA is arbitrarily rotated. If an image presents a large $\beta_2$, it is invariably either a MAD or a failing SANE. 

The phase angle of the $\beta_2$ coefficient trends with the magnitude of the black hole spin $\bhspin$. As can be seen in \autoref{fig:m2_spin}, increased $\left|\bhspin\right|$ correlates with smaller $\angle(\beta_2)$ and thus a more radially directed EVPA. A coherent external Faraday screen would uniformly rotate the EVPA map, preventing inference of spin if the rotation measure is not known.

The systematically negative phase of the $\beta_2$ coefficients is due to an imaging choice. In order to reproduce the ring asymmetry present in the EHT observation, the emitting fluid must be moving clockwise about the hole. This sets a preferred orientation for the black hole angular momentum vector and consequently also the magnetic field. Since EVPA traces magnetic field orientation, this choice preferences right-handed EVPA fields.

\section{Discussion}
\label{sec:discussion}

We have defined a decomposition of the linear polarization field into coefficients $\beta_m$ corresponding to ordered variation with respect to azimuthal angle in the image plane. The $\beta_2$ coefficient quantitatively identifies rotationally symmetric polarized image structure. The structural sensitivity of this decomposition provides additional constraints beyond measurements of the fractional polarization.
We have applied this decomposition procedure to a subset of the image library used by the EHT in their Stokes $I$ analysis of M87 and
identified a structure in the distribution of $\beta_2$ coefficient magnitudes that enables strong discrimination between the MAD and SANE accretion states.

Among images that obey observational and self-consistency criteria, only snapshots from MAD simulations are found to have $\left|\beta_2\right| > 0.2$ at 20 $\mu$as resolution. Further, we find that if intrinsic EVPA phase information can be recovered from the data, then the phase of the $\beta_2$ coefficient trends with black hole spin where increased spin correlates with a more radial EVPA. \added{This inference is possible even in the presence of an external Faraday screen if the rotation measure is well-constrained. The $\beta_2$ phase may be sufficient for discrimination of MAD and SANE even if the magnitude is low, as there are no passing SANE models in the negative real and imaginary quadrant of the $\beta_2$ distribution.} We also examined the signatures of rotationally symmetric polarization in interferometric visibilities and found that data at the current EHT resolution are likely sensitive to the differences in MAD and SANE polarization fields.

\added{We have not subdivided our results by $R_{\rm high}$ because electron heating physics are not well constrained by prior EHT efforts, nor particularly well discriminated in the current parameter space by our decomposition. However, we might generally expect that due to the fixed flux constraint in \citet{PaperV}, scrambling of ordered polarization by internal Faraday rotation would increase rapidly with $R_{\rm high}$. Very high values of $R_{\rm high}$ may cause sufficient disorder to be ruled out by coherence measures such as our decomposition. Other prescriptions for plasma thermodynamics in the accretion flow must be examined before the patterns observed in this study can be applied to real data.}

Although it is possible to analytically recover the position of the ring center in simulated images of black holes, we are limited to algorithmic centering procedures for analyses of real data. A polarimetric image of a centered ring with unit flux density given by $P(\varphi) = e^{i \left(2 \varphi + \delta\right)}$, as in Figure~\ref{fig:beta_2_ringplot},
will have $\left|\beta_2\right| = 1$. In ideal cases, \texttt{rex} can correctly identify the image center, enabling accurate computation of the decomposition coefficients. 
 
In more general cases, the turbulence in the underlying plasma flow can drastically affect the apparent shape and structure of the observed ring. These discrepancies, along with those due to undersampling in the image domain, can result in the \texttt{rex} procedure inaccurately identifying the true image center.
In orders of the absolute centering error $b$ divided by the ring diameter, $b/d$, the magnitude of the $\beta_2$ coefficient goes like the zeroth Bessel function of the first kind 
$\left|\beta_2\right| \approx J_0 \left( 4 \left| b/d\right|\right) \approx 1 - 4 \left( b / d \right)^2$.

For rings of finite width $w$, the correction term gains a factor of $\log\left( w / d \right)$. 
Note, however, that interferometric visibility amplitudes are invariant under centroid shifts, so the \added{amplitude} structures in \autoref{fig:mad_v_sane_fourier} persist regardless of image centering choices \citep{TMS}.

For Stokes $I$, the analysis in \citetalias{PaperV} found full image correlation times to be of order $50$ M, or approximately two weeks for M87. In contrast, we find that coherence in the $m=2$ mode can be significantly longer. Over the full image library, $m=2$ correlation times ranged from $50$ M to nearly $1000$ M. The increased coherence results primarily from the blurring and averaging procedures built into the coefficient calculation. Decoherence in the full images is generally driven by hot, isolated fluid features that evolve over a dynamical time. Because these features are localized in azimuth, they do not appear as time dependent features in the $m=2$ coefficient and thus $\beta_2$ retains coherence on longer timescales. 

\begin{figure}
    \centering
    \includegraphics[width=0.48\textwidth]{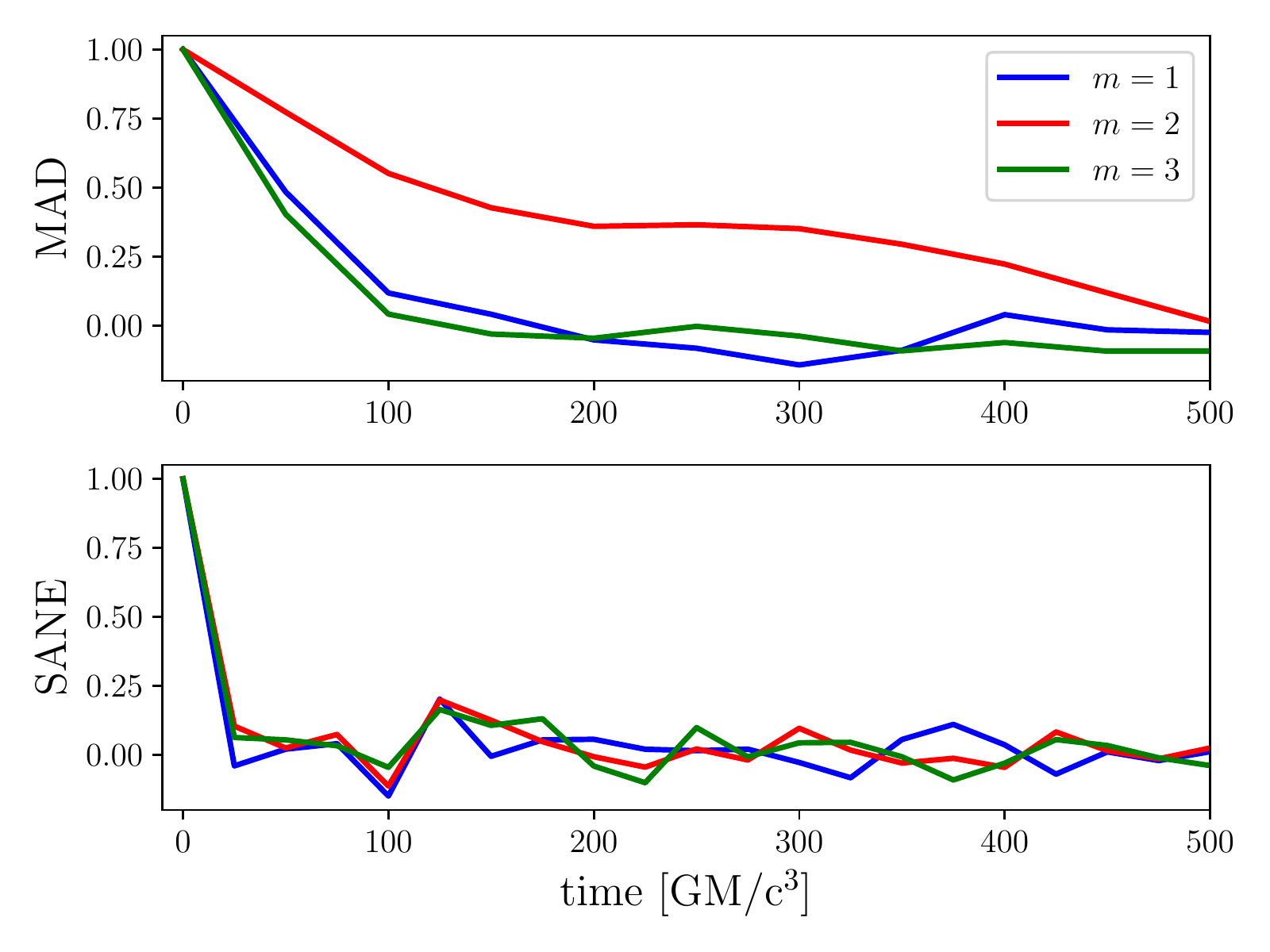}
    \caption{Autocorrelation function for several $\beta_m$ coefficients for an example MAD model and an example SANE model. In these models, MAD correlation times are longer than SANE correlation times. SANE correlation times may not be resolved by the image cadence.
    }
    \label{fig:madvsane_correlations}
\end{figure}

We also found that MAD correlation times are characteristically longer than SANE ones. This is reasonable: because the orientation of the EVPA is connected to the magnetic field structure and because magnetic fields are both stronger and more structured in MAD disks, EVPA orientation should be steadier in MAD disks. Our computation of correlation times is limited by image cadence, and because our cadence is at times of order the calculated coherence time, we have not attempted to provide a more detailed quantification of the difference between SANEs and MADs by this measure. \autoref{fig:madvsane_correlations} shows the autocorrelation function for both a selected MAD and a selected SANE model.
In the case of M87, we expect that images taken at intervals of a year or greater should provide independent realizations of the source and thus improve the statistical accuracy of any probability-based parameter discrimination.

The image library used for this analysis was generated using the fast light approximation in which it is assumed that the time it takes for light to travel through the computational domain is small compared to the timescales on which the fluid properties change. Our analysis is largely insensitive to this choice, especially because the inclination at which we view M87 orients the lines of sight perpendicular to the bulk motion of fluid features. 

Aside from the question of fast versus slow light, the values of the decomposition coefficients are geometrically dependent on the orientation at which we view M87. The nearly face-on view of M87 leads to an inherent symmetry which begets the relative strength of the $m=2$ mode.

Were the same analysis applied to images of black holes viewed edge on at large inclination, the symmetry would be \added{at least partially} broken. \added{Though the underlying magnetic field structures would still be imprinted on the polarized image, the image structure would no longer be as geometrically simple or convenient to describe along one azimuthal angle due to large foreground features that cross the shadow \citep[see, e.g.][]{Chael_2018sgra}. However, lensing guarantees the presence of flux in a ring-like structure at the shadow edge even if the emitting region is oriented edge-on with respect to the observer. In this case, a modeling procedure that separates ring-like emission from other structures may enable use of our decomposition.} The difference between fast and slow light analyses might also be more pronounced in analyses of black holes viewed closer to edge-on. \added{The presence of a tilted disk might create similar complications to larger inclinations, as the assumption of a nearly face-on viewing geometry would not hold for the emitting material even if true for the black hole spin vector or jet themselves. 

Also of interest are the recent observations of a flaring structure at the galactic center that seem to indicate face-on motion of a ``hot spot'' \citep{Gravity_2018_orbit}. The polarimetric variation found in this observation shows rotation of the polarization vector with period comparable to the hotspot orbital period, which we may associate with $|m| = 1$ variation. \autoref{fig:complex_all} shows that $\beta_1$ power can at times exceed $\beta_2$ power even in MAD models, while \autoref{fig:madvsane_correlations} indicates that $\beta_1$ is less stable in time than $\beta_2$. We then may expect that, despite the field symmetries in the steady flow, flaring structures could briefly excite large $m=1$ power due to light bending effects \citep[see appendix D of ][]{Gravity_2018_orbit}.}

Although our analysis was performed entirely within the image domain, images from observations are reconstructions from data \deleted{products that live} in the Fourier plane. Because assumptions about image structure affect the reconstruction procedure \citepalias[for a review, see][]{PaperIV}, an analysis of polarimetric imaging output based on synthetic data would help ensure that neither imaging choices nor systematic errors due to baseline coverage nor problems in leakage calibration dominate the signatures of rotational symmetry that our analysis identifies. Although examination of polarimetric imaging systematics is beyond the scope of this work, such an investigation is necessary before the parameter discrimination we describe can be applied to real data.

\acknowledgments

The authors wish to thank Michael Johnson and Charles Gammie for their insight, continual support, and for their many helpful comments that greatly improved the text. \added{We gratefully acknowledge Ramesh Narayan, Jason Dexter, Monika Moscibrodzka, Andrew Chael, and Alejandra Jim\'enez-Rosales for many illuminating discussions related to this work.} The authors also wish to thank Jae-Young Kim for his thorough review of the article. \added{Finally, we thank the referee for their careful and thoughtful feedback on our manuscript.}

DCMP was supported by NSF grant AST-1716536 and the Gordon and Betty Moore Foundation grant GBMF-5278. This work was supported by the Black Hole Initiative at Harvard University, which is funded by grants from the John Templeton Foundation and the Gordon and Betty Moore Foundation to Harvard University. GNW was supported by NSF grant AST 17-16327 and by the US Department of Energy through Los Alamos National Laboratory. Los Alamos National Laboratory is operated by Triad National Security, LLC, for the National Nuclear Security Administration of the US Department of Energy (Contract No.~89233218CNA000001). BSP was supported by NSF grant PIRE 17-43747. This work is authorized for unlimited release under LA-UR-19-31256.

The authors acknowledge the Texas Advanced Computing Center (TACC) at The University of Texas at Austin for providing HPC resources that have contributed to the research results reported within this paper. 
\software{eht-imaging \citep{Chael_closure},
Numpy \citep{numpy},
Matplotlib \citep{matplotlib},
GNU Parallel \citep{Tange2011a}}

\pagebreak
\bibliographystyle{yahapj}
\bibliography{references}

\end{document}